\documentclass[journal abbreviation]{copernicus}

\begin{document}

\title{COLLISIONLESS MAGNETIC RECONNECTION IN  A PLASMOID CHAIN}

\author[1,2]{Stefano Markidis}
\author[1]{Pierre Henri}
\author[1]{Giovanni Lapenta}
\author[1]{Andrey Divin}
\author[3]{Martin V. Goldman}
\author[3]{David Newman}
\author[4]{Stefan Eriksson}

\affil[1]{Centrum voor Plasma-Astrofysica, Departement Wiskunde, Katholieke Universiteit Leuven, Belgium}
\affil[2]{PDC Center for High Performance Computing, KTH Royal Institute of Technology, Stockholm, Sweden}
\affil[3]{Department of Physics and CIPS, University of Colorado, Boulder, Colorado 80309, USA}
\affil[4]{Laboratory for Atmospheric and Space Physics, University of Colorado, Boulder, Colorado 80309, USA}
%\affil[]{ADDRESS}

%% The [] brackets identify the author to the corresponding affiliation, 1, 2, 3, etc. should be inserted.

\runningtitle{PLASMOID CHAIN RECONNECTION}

\runningauthor{S.Markidis, P.Henri, G.Lapenta, A.Divin, M.V.Goldman, D.Newman, S.Eriksson }

\correspondence{Stefano Markidis\\ (markidis@pdc.kth.se)}

\received{}
\pubdiscuss{} %% only important for two-stage journals
\revised{}
\accepted{}
\published{}

%% These dates will be inserted by the Publication Production Office during the typesetting process.

\firstpage{1}

\maketitle

\begin{abstract}
The kinetic features of plasmoid chain formation and evolution are investigated by two dimensional Particle-in-Cell simulations. Magnetic reconnection  is initiated in multiple $X$~points by the tearing instability. Plasmoids form and grow in size by continuously coalescing. Each chain plasmoid exhibits a strong out-of plane core magnetic field and an out-of-plane electron current that drives the coalescing process. The disappearance of the $X$~points in the coalescence process are due to anti-reconnection, a magnetic reconnection where the plasma inflow and outflow are reversed with respect to the original reconnection flow pattern. Anti-reconnection is characterized by the Hall magnetic field quadrupole signature. Two new kinetic features, not reported by previous studies of plasmoid chain evolution, are here revealed. First, intense electric fields develop in-plane normally to the separatrices and drive the ion dynamics in the plasmoids. Second, several bipolar electric field structures are localized in proximity of the plasmoid chain. The analysis of the electron distribution function and phase space reveals the presence of counter-streaming electron beams, unstable to the two stream instability, and phase space electron holes along the reconnection separatrices. 
\end{abstract}

%% only used for copernicus2.cls
%\abstract{
% TEXT
  %}

\introduction
%% \introduction[modified heading if necessary]
Magnetic reconnection is believed to be the key engine to conversion of magnetic energy into heating and acceleration of space and astrophysical plasmas. Energy is typically stored as magnetic energy, then suddenly released released in form of violent plasma jets and non-thermal particle acceleration. Magnetic reconnection is a basic plasma physics phenomenon triggering space weather events such as magnetospheric substorms or solar flares. Because of the ubiquity of this phenomenon in nature and because it is at the base of macroscopic solar physics and magnetospheric events, many theoretical, computational and experimental studies have been devoted to explain the origin and the dynamics of magnetic reconnection. However, one main question has still no certain answer. It is unclear why magnetic reconnection in nature proceeds on faster time scale than the resistive diffusion timescale predicted by the Sweet-Parker model in the magnetohydrodynamics (MHD) framework. It has been shown the all simulation models including the Hall physics, based the decoupling of electrons and ions at ion skin depth scale, enable reconnection to occur over faster time period \citep{GEM:2001}. It has been proposed recently that reconnection occurring in multiple points and secondary plasmoid generation can lead to fast reconnection in the MHD context  also via stochastic plasmoid chain formation \citep{Lazarian:1999, Kowal:2009,Lapenta:2008,bhattacharjee:2009,Uzdensky:2010}. The present paper studies the combination of these two mechanisms to achieve fast reconnection: a kinetic model of plasma, where Hall physics is a built-in feature, is used to study plasmoid chain magnetic reconnection.

The goal of this paper is to study the formation and evolution of plasmoid chain structures in the kinetic framework by two dimensional Particle-in-Cell simulations. Plasmoids are high density structures forming from the outflow plasma accelerated in magnetic reconnection \citep{Hones:1977}. A closed magnetic loop develops in correspondence of the plasmoids, this is why the expression {\em magnetic island} is often used instead of plasmoid. The boundary layer between the plasmoid and the inflow plasma is called {\em separatrix}. If magnetic reconnection occurs in multiple points, the outflow plasmas from adjacent reconnection sites form multiple plasmoids, organized as beads in a chain. Plasmoid chains are highly dynamic configurations: plasmoids can coalescence forming larger plasmoids \citep{Bhattacharjee:1983}, bounce \citep{Intrator:2009,Karimabadi:2011} and eject smaller plasmoids \citep{Daughton:2009}. Observational studies of solar and magnetospheric plasmas confirmed the presence of  plasmoids (flux ropes in three dimensions) in magnetospheric and solar environments. The plasmoid chains have been observed in sun corona \citep{Bemporad:2008} and in Earth magnetopause and magnetotail \citep{Lin:2008, Khotyaintsev:2010}.

The formation and evolution of plasmoid chain are a topic of many recent theoretical and computational investigations. Early simulations studied the occurrence of plasmoid chain by MHD simulations \citep{Biskamp:1986,Malara:1992}. Theoretical models and simulations showed how the occurrence of secondary plasmoids affects magnetic reconnection and and how fast reconnection in low resistivity plasmas can be achieved in the MHD framework via stochastic magnetic reconnection \citep{Lazarian:1999, Kowal:2009, Loureiro:2005, Loureiro:2007, Lapenta:2008,bhattacharjee:2009,Cassak:2009, Uzdensky:2010}. The realization of fast magnetic reconnection has also been investigated by Hall MHD simulations \citep{Baalrud:2011}. Collisionless Particle-in-Cell simulations \citep{Tajima:1987,Pritchett:2008,OkaAPJ:2010, OkaJGR:2010,Tanaka:2010} studied the dynamics of plasmoid chain. The focus of previous Particle-in-Cell studies was primarily on the mechanisms that lead electron acceleration to relativistic energies. Differently from previous works on collisionless plasmoid chain, the goal of this paper is to study the general properties of magnetic reconnection in a plasmoid chain in the kinetic framework, giving a broad picture of the plasmoid formation and evolution. Two new results are here reported. First, an intense electrostatic activity driving the ion dynamics is detected along the reconnection separatrices in the plasmoid chain. Second, counter-streaming electron beams along the separatrices are unstable to the two-stream instability and electron phase space holes appear as a result of this instability.
 
The paper is organized as follows. First, the simulation model is presented. Second, the formation and evolution of plasmoid chain, the magnetic and electric field configurations are analyzed; a study of ion and electron distribution functions and phase space are presented to unveil possible acceleration mechanisms and the presence of micro-instabilities in proximity of the plasmoid chains. Third, the simulation results are discussed and compared with those of previous works. Finally, the results are summarized and future work is outlined in the conclusion section.

\section{Simulation set-up}
Particle-in-Cell simulations are carried out in a two dimensional space, the $X-Y$ plane with $Z$ as ignorable direction, with three components for particle velocities and field components. The $x$, $y$ coordinates correspond to the $-x_{GSM}$ (Earth-Sun direction in magnetotail), $z_{GSM}$ (North-South direction in magnetotail) coordinates in the Geocentric Solar Magnetospheric (GSM) system. 

A double current sheet with reversing magnetic field in the $X$ direction is initialized as:
 \begin{equation}
B_x/B_0= \tanh(\frac{y - L_y/4}{\lambda}) - \tanh(\frac{y - 3L_y/4}{\lambda}) - 1.
\end{equation}
A uniform guide field is present along the out-of-plane $Z$ direction and equal to $B_0/2$.
The initial density profile is a double Harris current sheet with a superimposed background plasma with density $n_b$:
\begin{equation}
%n(y) = 0.2 n_0 (\cosh^{-2}(\frac{y - L_y/4}{\lambda}) +  \cosh^{-2}(\frac{y - 3L_y/4}{\lambda})),
n(y) = 0.2 \ n_0 \ \Big( \cosh^{-2} \big( \frac{y - L_y/4}{\lambda} \big) +  \cosh^{-2} \big( \frac{y - 3L_y/4}{\lambda} \big) \Big),
\end{equation}
where $n_0$ is the reference density and $n_{b} =  0.2\ n_0$; $L_x = 200\ d_i$, $L_y = 30 \ d_i$ and $\lambda = 0.5\ d_i$ are the simulation box lengths in the $X$ and $Y$ directions and the half-width of the double current sheet respectively. The lengths are normalized to $d_i = c/\omega_{pi} $, the ion inertial length,  with $c$ the speed of light in vacuum and $\omega_{pi} = \sqrt{(4 \pi n_0e^2/m_i)}$. The ratio between the ion plasma and cyclotron frequencies is $\omega_{pi}/\Omega_{ci}= 1030.9$. The ion mass $m_i$ is taken $256$ larger than electron mass $m_e$ and $e$ is the elementary charge. The plasma is initialized with a Maxwellian velocity distribution. The electron thermal velocity is chosen $ v_{the}/c =(T_e/(m_e c^2))^{1/2} = 0.0045$. Ion temperature is $T_i = 5 \ T_e$. The simulation parameters are chosen to mimic the plasma characteristics in the solar wind current sheets as observed by Wind and Cluster spacecrafts \citep{Stefan}. In order to speed-up the occurrence of magnetic reconnection, instead of using a perturbation as in previous studies, the initial system configuration is chosen not in equilibrium. The initial current is 20\% of the current necessary to support consistently the initial magnetic field configuration. As result of this non-equilibrium, the plasma is initially accelerated toward the double current sheet to establish a current consistent with the initial magnetic field configuration, accelerating the reconnection initiation.

The simulation time step is $\omega_{pi}\Delta t = 0.3$. The grid is composed of $2560\times 384$ cells, resulting in a grid spacing $\Delta x  = \Delta y = 0.078 \ d_i$. In total $7\times 10^8$ computational particles are in use. Periodic boundary conditions for particles and fields are applied in the $X$ and $Y$ directions. Simulations are carried out with the parallel implicit Particle-in-Cell iPIC3D code \citep{Markidis:2010}. These simulations would be extremely difficult for be carried out with explicit Particle-in-Cell codes because of the size of the simulation box and the simulated time period. If compared with typical parameters in explicit Particle-in-Cell simulations, large time step, grid spacing and ion to electron mass ratio have been used. Numerical stability is retained, thanks of the implicit discretization in time of the Particle-in-Cell governing equations \citep{Lapenta:2006}. In fact, a time step $\Delta t = 0.3 \ \omega_{pi}^{-1} = 4.8 \ \omega_{pe}^{-1}  $ makes the explicit Particle-in-Cell method unstable in these parameters regime. Moreover, the grid spacing in use is approximately 277  Debye lengths. The possibility of using grid spacings that are much larger than Debye length still avoiding aliasing instabilities, such as the finite-grid instability, is a property of the implicit Particle-in-Cell. Moreover, an ion to electron mass ratio equal to 256 is here used, instead of 25 and 100 in previous studies with explicit Particle-in-Cell codes \citep{Pritchett:2008,OkaAPJ:2010, OkaJGR:2010,Tanaka:2010}. The effect of the implicit numerical scheme is to damp artificially the unresolved waves, such as the light and Langmuir waves \citep{brackbill-cohen-85,Markidis:2011}. The implicit Particle-in-Cell algorithm still correctly describes phenomena evolving on time scales that are larger than the time step. Magnetic reconnection occurs over time period of tens of $\Omega_{ci}^{-1}$, approximately ten thousand times the simulation time step, and the implicit numerical scheme does not alter its evolution.

\section{Plasmoid chain formation and evolution}
A tearing instability rapidly occurs out of particle noise and generates several plasmoids. The reconnected flux for the single plasmoid $\psi$ is calculated as the difference of the out-of-plane component of the vector potential $A_z$ at $X$ and $O$ points along the line $y=3/4 \ L_y= 22.5 \ d_i$ (center of the top current sheet). Figure~ 1 panel a  shows a stack plot of the out-of-plane component of the vector potential $A_z$, normalized using the background ion skin depth and the asymptotic magnetic field $B_0$. An $A_z$ minimum identifies an $X$~point (green and red colors in Figure~ 1 panel a, while a maximum (black color) indicates an $O$ point. The plot illustrates multiple plasmoids collapse, forming larger plasmoids. The reconnection rate is used to characterize the speed at which magnetic reconnection occurs. A separate analysis of the reconnection rate at different reconnection sites is required, because magnetic reconnection develops in a large number of points. In addition, the reconnection sites move in time, it is therefore necessary to identify the reconnection site and follow them during the simulation. The reconnection rate is calculated for each reconnection site, indicated with solid lines in Figure~ 1 panel a. The reconnection rate can be computed in two ways: (i) the time derivative of the reconnected flux $\psi$ (solid lines in Figure~1 panel b)  (ii) the reconnection electric field $E_z$ in the upstream region, normalized to the Alfv\'en velocity $V_A$ and $B_0$ (dashed lines in Figure~ 1 panel b). The two measures of reconnection rates are in agreement in each site. The reconnection rate evolves very similarly for all the $X$~points: it grows very rapidly until time $\Omega_{ci}t = 1.5$, when it reaches the 0.3 peak value, then decreases reaching the range $0.01- 0.08$ for different $X$~points. 

\begin{figure*}[ht!]
\vspace*{2mm}
\begin{center}
\includegraphics[width=14cm]{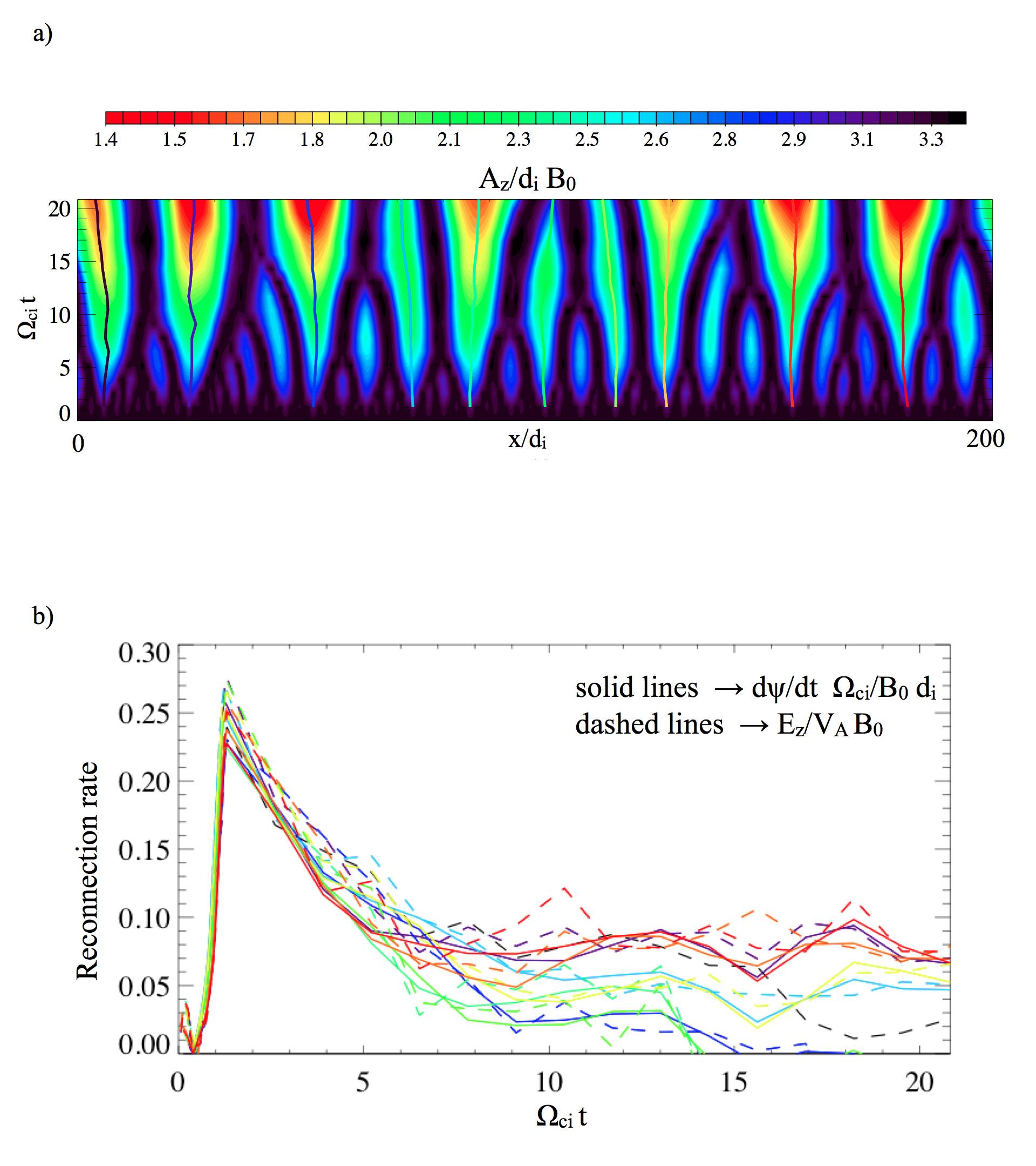}
\end{center}
\caption{The stack plot of the out-of-plane vector potential component $A_z$ along the line $y = 22.5 \ d_i$  is shown in panel a. The reconnection rate in panel b) is calculated as derivative in time of the reconnected flux $\psi$ (solid line) and as $E_z$ (dashed lines), for different $X$~points (colored lines in panel a. }
\end{figure*}

As result of magnetic reconnection, several high density structures are formed by the outflow plasmas of adjacent $X$~points (Figure~ 2, time $\Omega_{ci}t = 3.5$). The electron plasmoid density is approximately four times the background density. At time $\Omega_{ci}t = 3.5$, plasmoids are large $5 \ d_i$ and progressively grow by merging and forming larger plasmoids. Low density layers, called {\em cavities} \citep{Pritchett:2008}, form along the reconnection separatrices. During the simulation, the expulsion and formation of smaller plasmoids are not observed. Because secondary islands are not produced during magnetic reconnection, the plasmoids have approximately the same spatial size and very large plasmoids, such as {\em monster} plasmoids \citep{Uzdensky:2010}, do not appear. The hierarchical self-similar structure of the plasmoid-dominated current sheet is not detected in the Particle-in-Cell simulations \citep{Uzdensky:2010}.
\begin{figure*}[ht!]
\vspace*{2mm}
\begin{center}
\includegraphics[width=16cm]{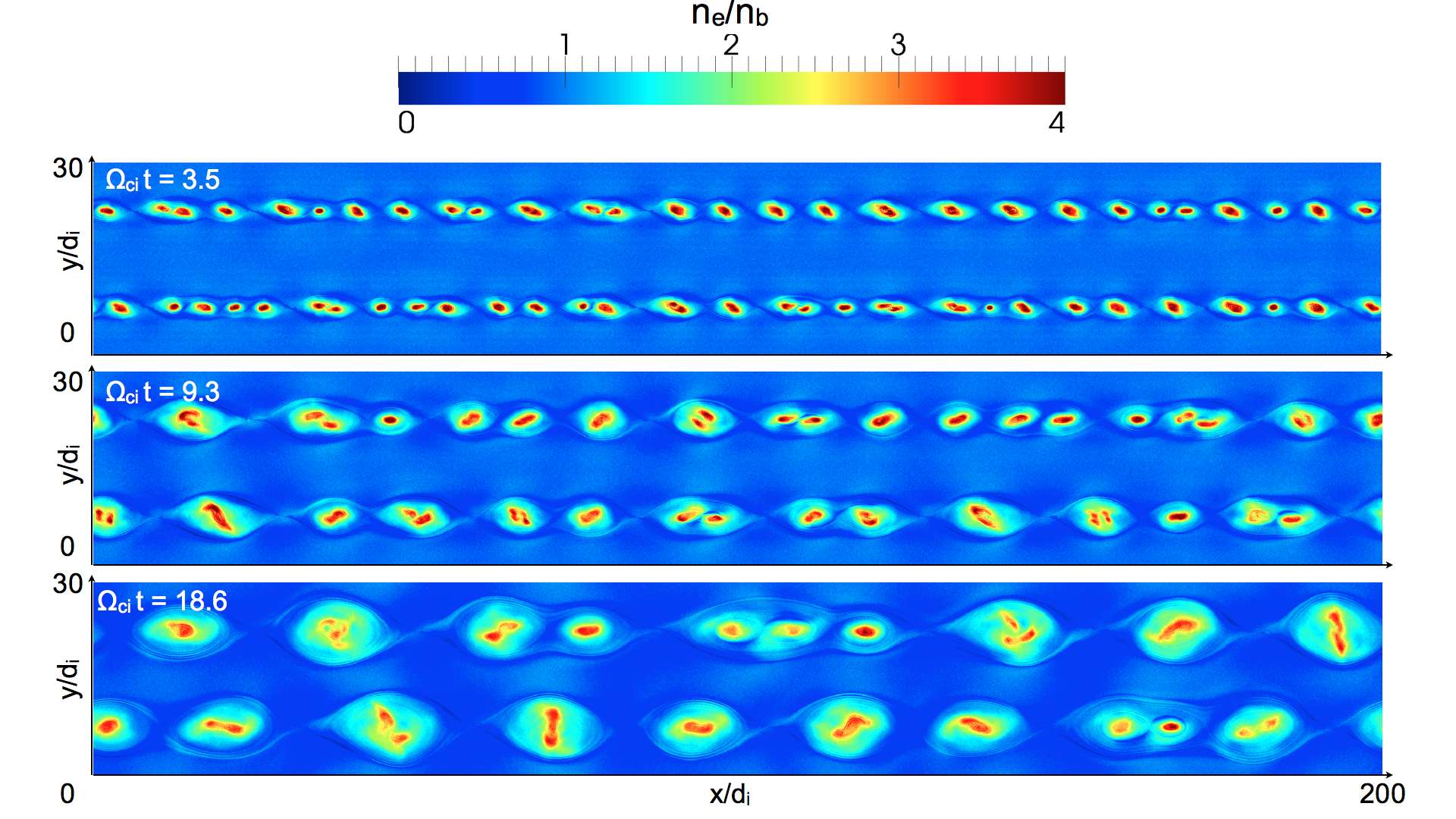}
\end{center}
\caption{Contour plots of the electron density, normalized to the background density $n_b$, at different times. The plot shows the formation and the progressive growth of relatively high density regions, the plasmoids, by coalescence.}
\end{figure*}

All the plasmoids develop a relatively strong out-of-plane core magnetic field, as shown in the $B_z$ contour plot in Figure~ 3. The core magnetic field is unipolar and in the same direction for the plasmoids of both current sheets. The intensity of $B_z$ magnetic field without the guide field is approximately $0.5 \ B_0$, that is the value of the guide field in the proposed simulation.
\begin{figure*}[!ht]
\vspace*{2mm}
\begin{center}
\includegraphics[width=16cm]{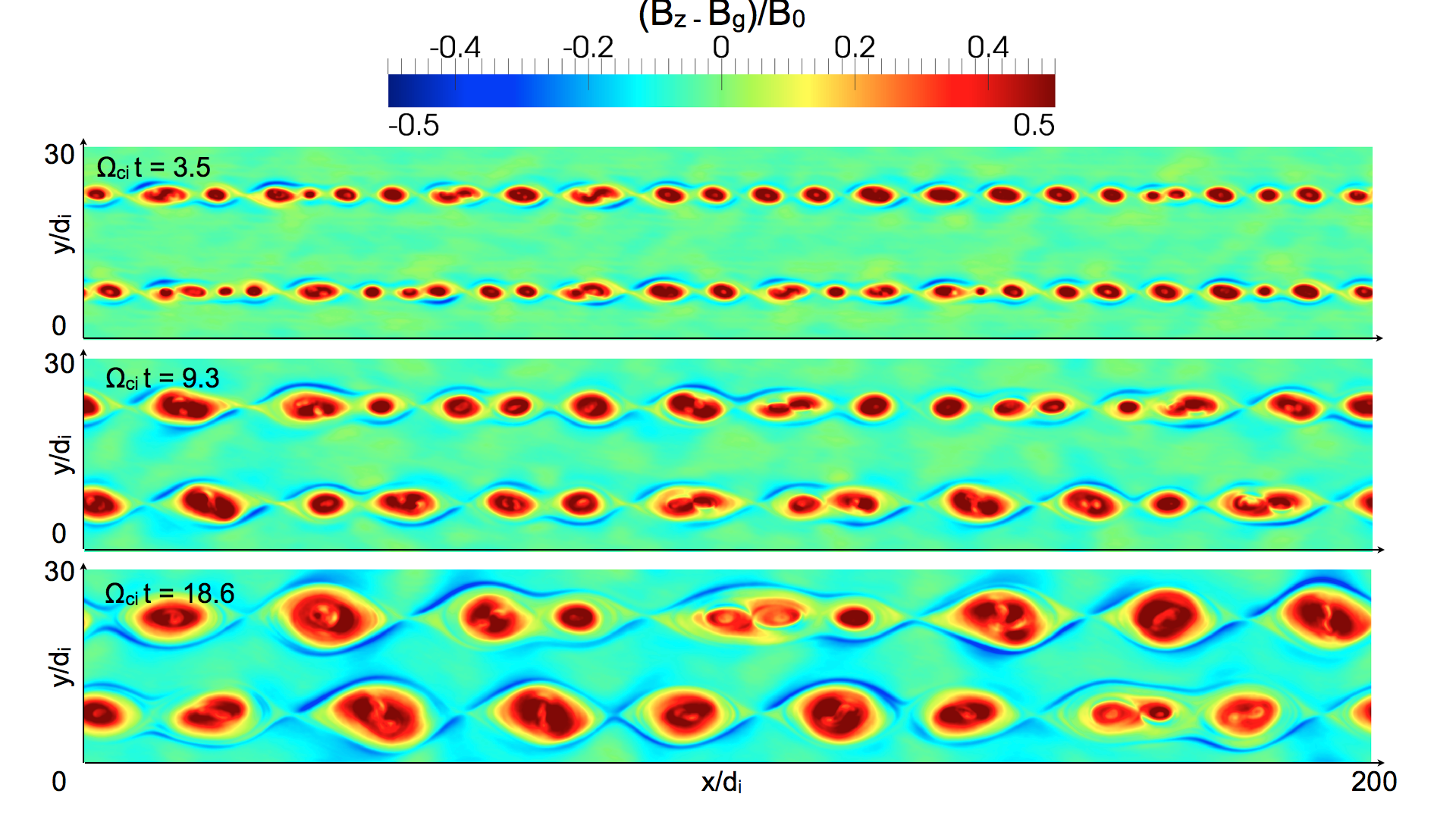}
\end{center}
\caption{Contour plots of the out-of-plane magnetic field without the guide field, $B_z - B_g$, normalized to $B_0$ at different times. Each magnetic island is characterized by an out-of-plane core magnetic field, whose value is approximately equal to the guide field $B_g = 0.5 B_0$.}
\end{figure*}
As noted in previous works \citep{Karimabadi:1999}, the core magnetic field is not uniform over the plasmoids, but it presents a complex pattern. The S-shaped $B_z$ regions, that were found in hybrid simulations of multiple $X$~points reconnection, are present also in the $B_z$ contour plot in Figure~ 4. The coalescence process drives the anti-reconnection \citep{OkaAPJ:2010,Tanaka:2010}, where magnetic reconnection inflow and outflow are reverted. The inflow is directed in the $X$ direction, while the outflow direction in anti-reconnection is along the $Y$ direction (white arrows in Figure~ 4). The signature of anti-reconnection is visible in the typical quadrupolar structure of the Hall magnetic field in the white dashed squares in Figure~ 4. The quadrupole Hall magnetic field is rotated $90 ^{\circ}$ with respect to the typical reconnection Hall magnetic field and generates the S-shaped regions, observed previously in multiple $X$~points reconnection hybrid simulations \citep{Karimabadi:1999}.
\begin{figure*}[!ht]
\vspace*{2mm}
\begin{center}
\includegraphics[width=16cm]{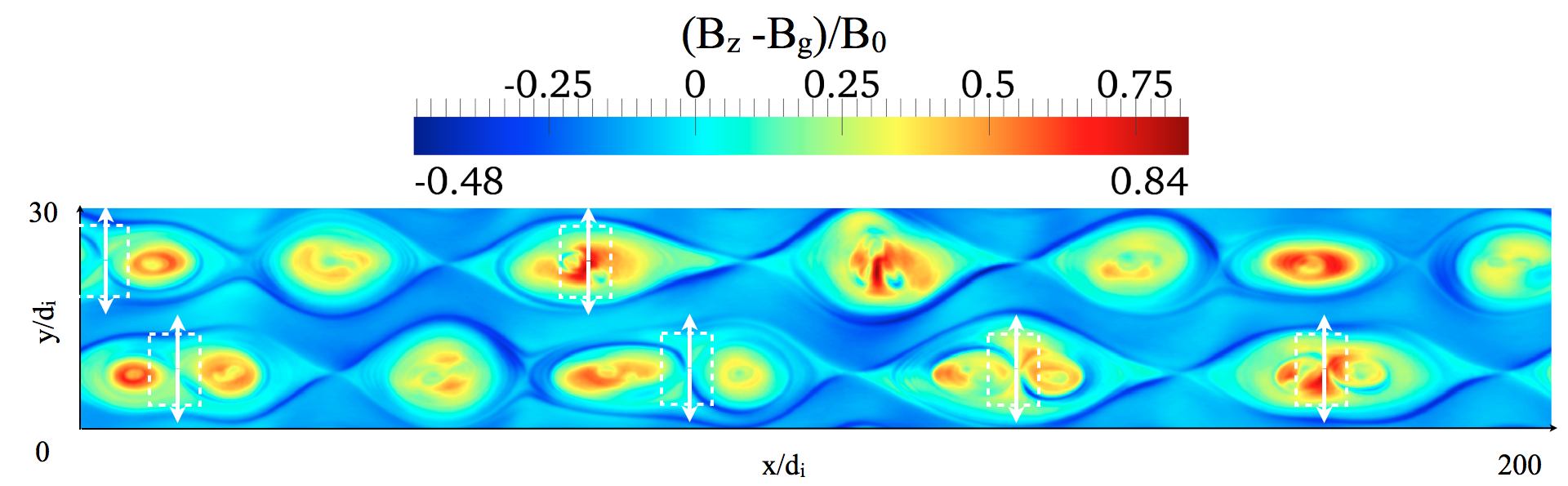}
\end{center}
\caption{Contour plots of the out-of-plane magnetic field without the guide field, $B_z - B_g$, normalized to $B_0$ at time $\Omega_{ci}t = 26.77$. The core magnetic field shows a complex pattern. The S-shaped regions are visible in the coalescing area. The quadrupole structure of the Hall magnetic field in the coalescence-driven anti-reconnection is enclosed in white dashed rectangles. The white arrows indicate the outflow direction in anti-reconnection.}
\end{figure*}

The out-of-plane current component of the current drives the coalescence process. In fact, plasmoids act as wires carrying current in the same direction attracting each other, coalescing one plasmoid into another one. The total current is supported only by the electrons: the intensity of electron current is seven times higher than the ion current, developing in the opposite direction (Figure~ 5). This result is in agreement with the three-dimensional kinetic simulations of flux ropes, where the axial current is supported only by electrons \citep{MarkidisEPS}. 

\begin{figure*}[!ht]
\vspace*{2mm}
\begin{center} with
\includegraphics[width=16cm]{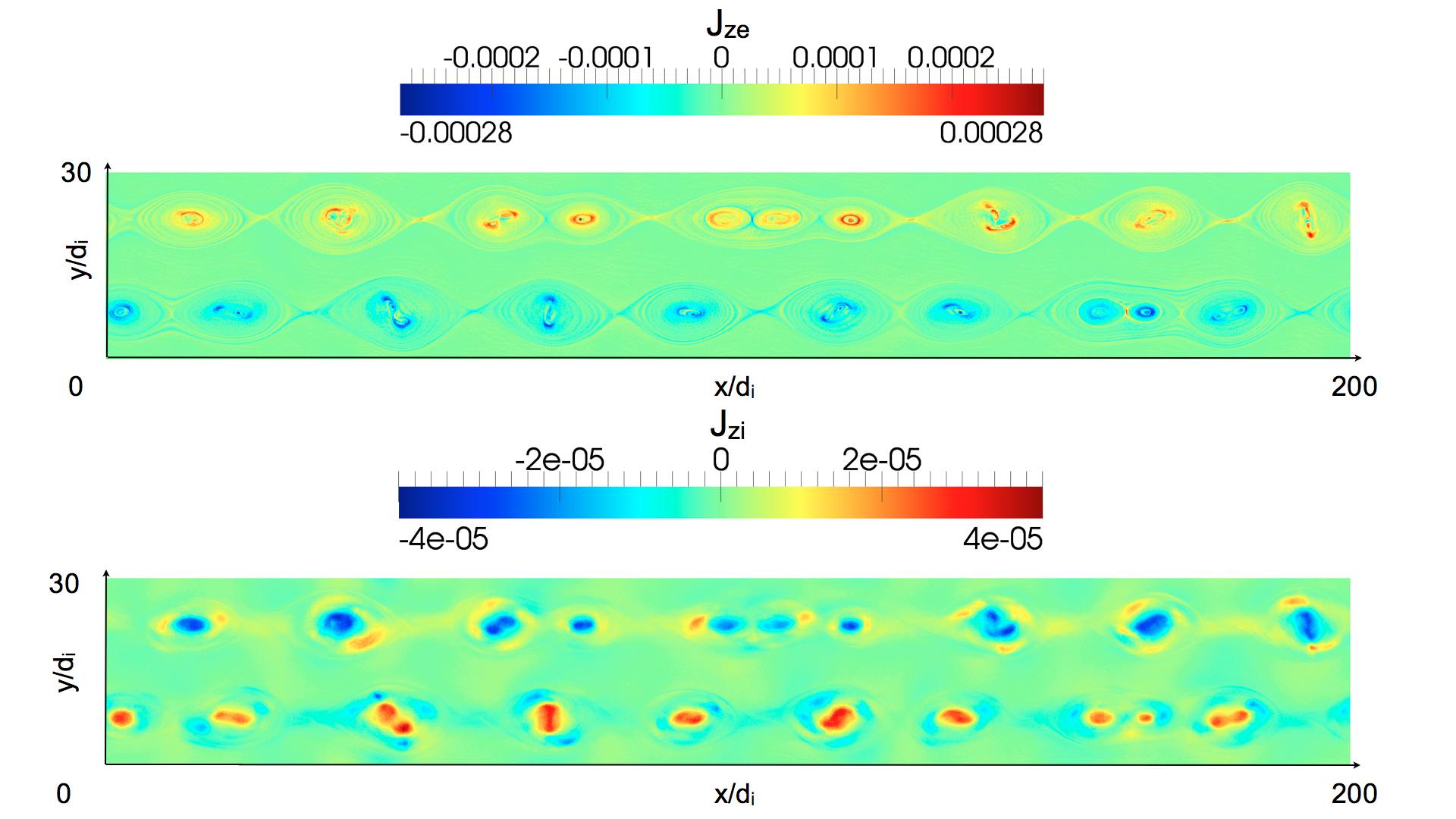}
\end{center}
\caption{Contour plot of the out-of plane electron and ion currents, $J_{ze}$ and $J_{zi}$, normalized to $e n_0 c$, at time $\Omega_{ci}t = 18.62$. The total current is supported only by the electrons while the the ions generate a current in the opposite direction.}
\end{figure*}

The most intense electric fields are in-plane and localized along the separatrices or in the plasmoids, as shown in Figure~ 6. The maximum value of the electric field is $\sim 2 \ V_A B_0 /c$, that is approximately 10 times the peak reconnection rate.
\begin{figure*}[!ht]
\vspace*{2mm}
\begin{center}
\includegraphics[width=16cm]{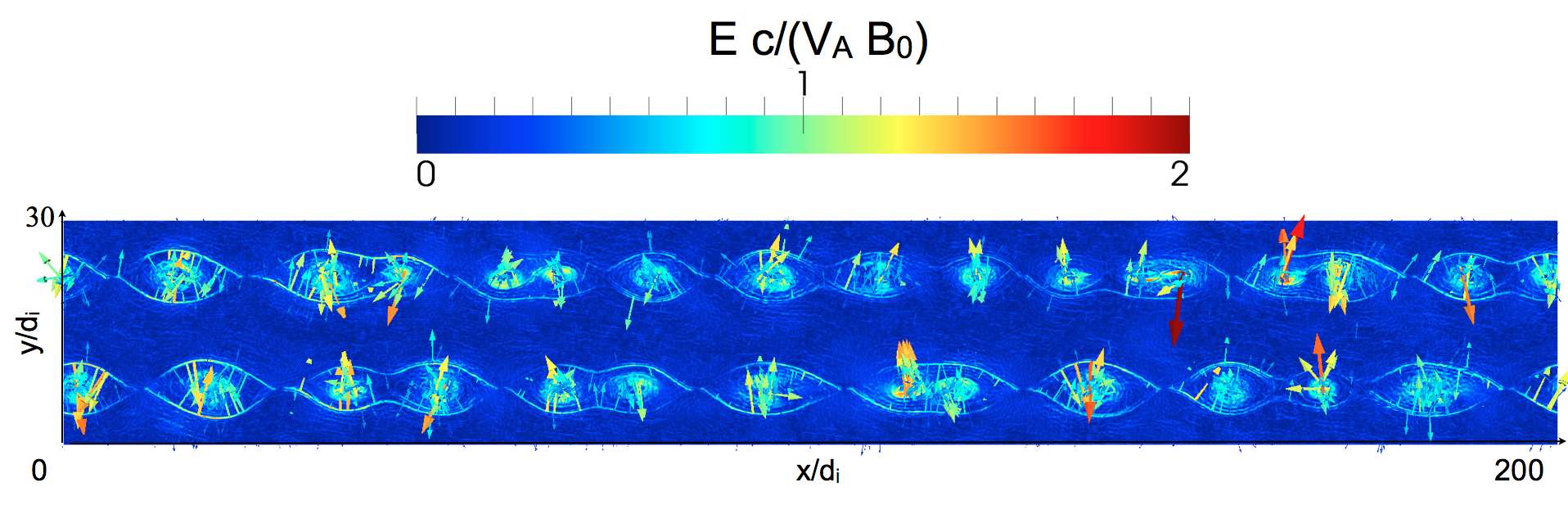}
\end{center}
\caption{Quiver plot and contour plot of the electric field, normalized to the reconnection electric field $V_aB_0/c$, at time $\Omega_{ci} t = 18.62$. The electric field is mainly in-plane and is localized around and inside the magnetic islands.}
\end{figure*}

The electric field configuration in a plasmoid region is shown in detail in an enlargement of an area occupied by a plasmoid at time $\Omega_{ci} t = 18.62$ in Figure~7. A strong inward electric field is present in its center. Along the separatrices, the electric field is normal to the separatrices (black line in panel b): it is directed inward inside the separatrix, while it is outward just outside the separatrix. Thus, ions in the plasmoid are accelerated inward, while electrons are accelerated outward and reflected back in the plasmoid. The ion bulk flow in a plasmoid is presented in Figure~ 7 panel b. In guide field reconnection, the reconnection jets are not completely directed along the $X$ direction, but their directions are tilted by the guide field \citep{Goldman:2011}. Thus, the reconnection jets from different $X$~points do not run head-to-head, but stream one over the other. The strong ion bulk flow is localized inside the plasmoid. An analysis of the ion phase space $v_x$ and $v_y$ vs $x$ in a $0.234 \ d_i$ thick region along $y = 22.5 \ d_i$ (blue dashed line in panel b) shows that ions are accelerated towards the plasmoid center, and in the $-Y$ and $Y$ directions for small and large $X$ values respectively. This combined motion generates a rotation of the ion bulk flow in the counterclockwise direction. Outside the separatrix line, a weak ion flow rotates in the opposite direction. This ion bulk velocity system is consistent with the ion current pattern, presented in Figure~ 5 panel b. 
\begin{figure*}[!ht]
\vspace*{2mm}
\begin{center}
\includegraphics[width=12cm]{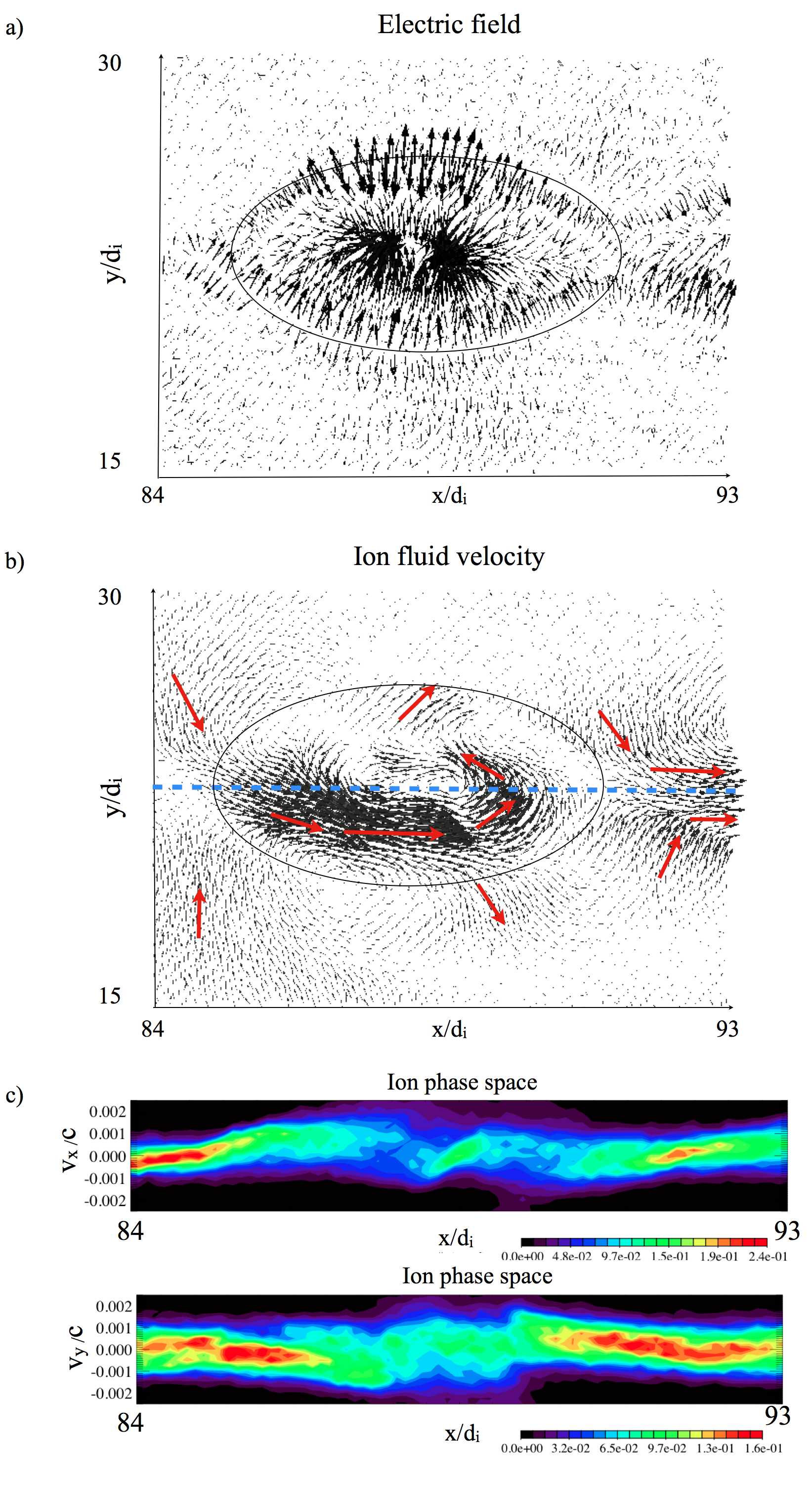}
\end{center}
\caption{Enlargement of a simulation region, $x = 84 - 93, y = 15 - 30$ enclosing a plasmoid at time $\Omega_{ci} t = 18.62$. A quiver plot of the electric field and ion bulk flow are shown in panels a and b. The length of the arrows indicates the intensity. The red arrows shows the direction of the ion bulk flow. The phase-space plots, $v_x$ and $v_y$ vs x at $y = 22.5$ (blue dashed line in panel b) is shown panel c.}
\end{figure*}

In magnetic reconnection studies, the bipolar electric field signatures are regarded as flag of electron micro-instabilities \citep{LapentaGRL:2011}. An investigation of the parallel electric field $E_{//} = \mathbf{E}\cdot \mathbf{B}/|B|$ in Figure~8 reveals that there are several bipolar parallel electric field structures inside the plasmoids and along the separatrices. These signatures have maximum and minimum values $\pm 0.3 B_0V_A/c$, approximately equal to peak reconnection electric field. 
\begin{figure*}[!ht]
\vspace*{2mm}
\begin{center}
\includegraphics[width=12cm]{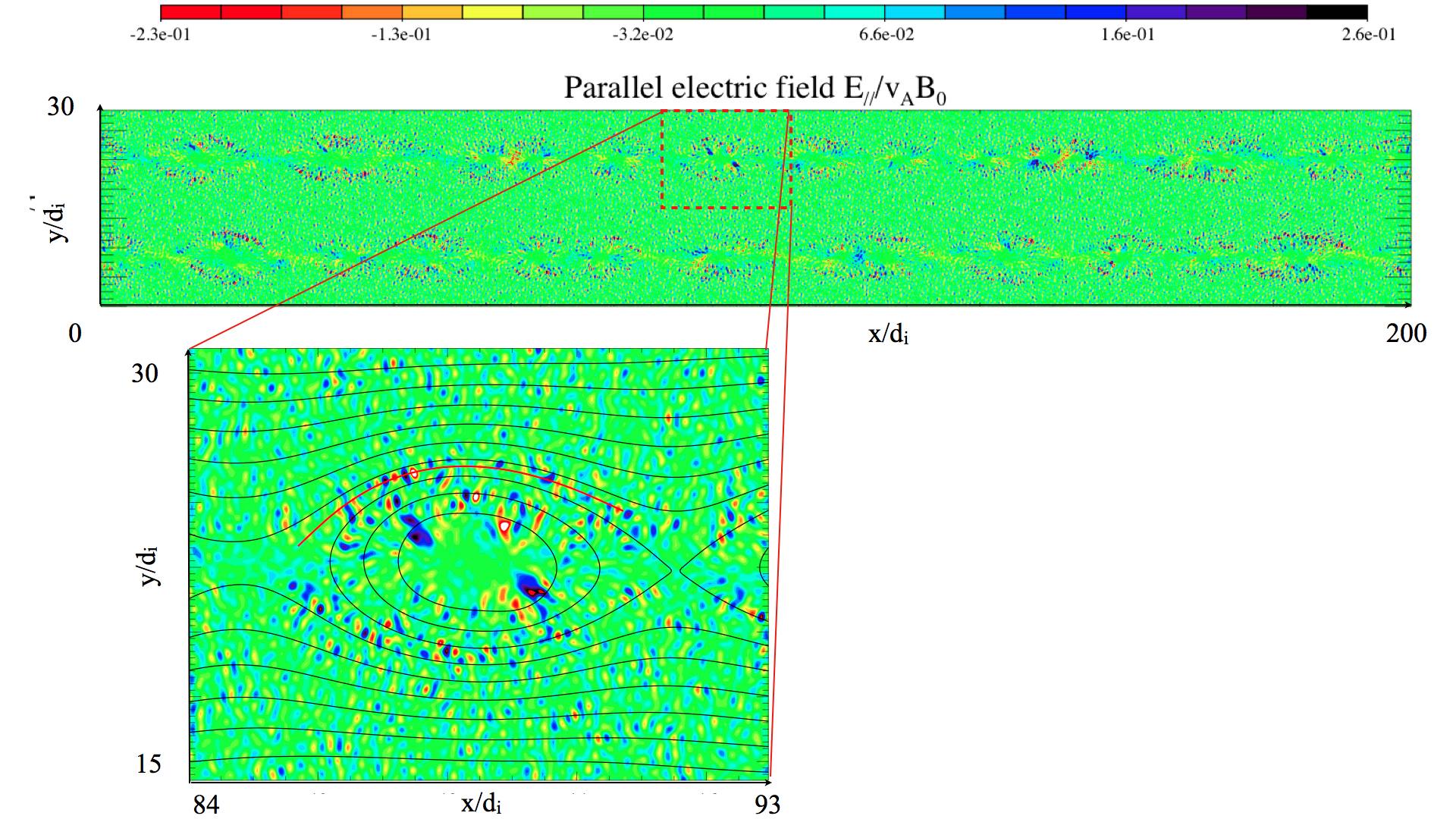}
\end{center}
\caption{Contour plot of the parallel electric field $E_{//} = \mathbf{E}\cdot \mathbf{B}/|B|$, normalized to the reconnection electric field $B_0V_A/c$ with superimposed magnetic field lines in black at time $\Omega_{ci} t = 18.62$ and enlargement of of a simulation region, $x = 84 - 93, y = 15 - 30$ enclosing a plasmoid.  Bipolar parallel electric field signatures are localized in proximity of the magnetic separatrices. }
\end{figure*}

An enlargement of the same region presented in Figure~7 shows that bipolar electric field structures are aligned along the magnetic field lines, suggesting the presence of a beam instability. The electron distribution function is analyzed to verify this hypothesis. The electron distribution functions, $f_e(v_x,v_y)$, in four different locations are presented in Figure~ 9 panel a. The four different regions, black boxes in the  central subpanel, have been chosen along the separatrix (red line in the $E_{//}$ contour plot in the central subpanel), covering a $ 0.234 \ d_i \times 0.234 \ d_i $ wide area. A comparative analysis of the four electron distribution functions shows the presence of counter-streaming electron beams along the magnetic field. In subpanel I, the  electron distribution is taken in proximity of an $X$~point and is approximately Maxwellian. In subpanel II, the distribution function shows an electron core population at positive $v_x$ and $v_y$, a beam moving from left to right along the magnetic field direction (red line in the subpanels). An electron beam at negative $v_x$ and $v_y$ is present also.  Sub-panels III and IV confirm the presence of two beams, one at positive $v_x$ and $v_y$ and one at negative $v_x$ and $v_y$, moving in opposite directions.

Figure~ 9 panel b shows the electron phase space along a separatrix, $x_{//}$ vs $v_{//}$ (curvilinear coordinate along the magnetic field line vs the parallel velocity). For high and low $x_{//}$ values, two electron beams are visible as two electron populations moving at opposite speeds. The interaction of two beams results in the formation of phase space electron holes, in white dashed lines in Figure~ 9 panel b. There is clear indication that electron holes are generated by two-stream instabilities of the two electron beams along the separatrices.
\begin{figure*}[!ht]
\vspace*{2mm}
\begin{center}
\includegraphics[width=12cm]{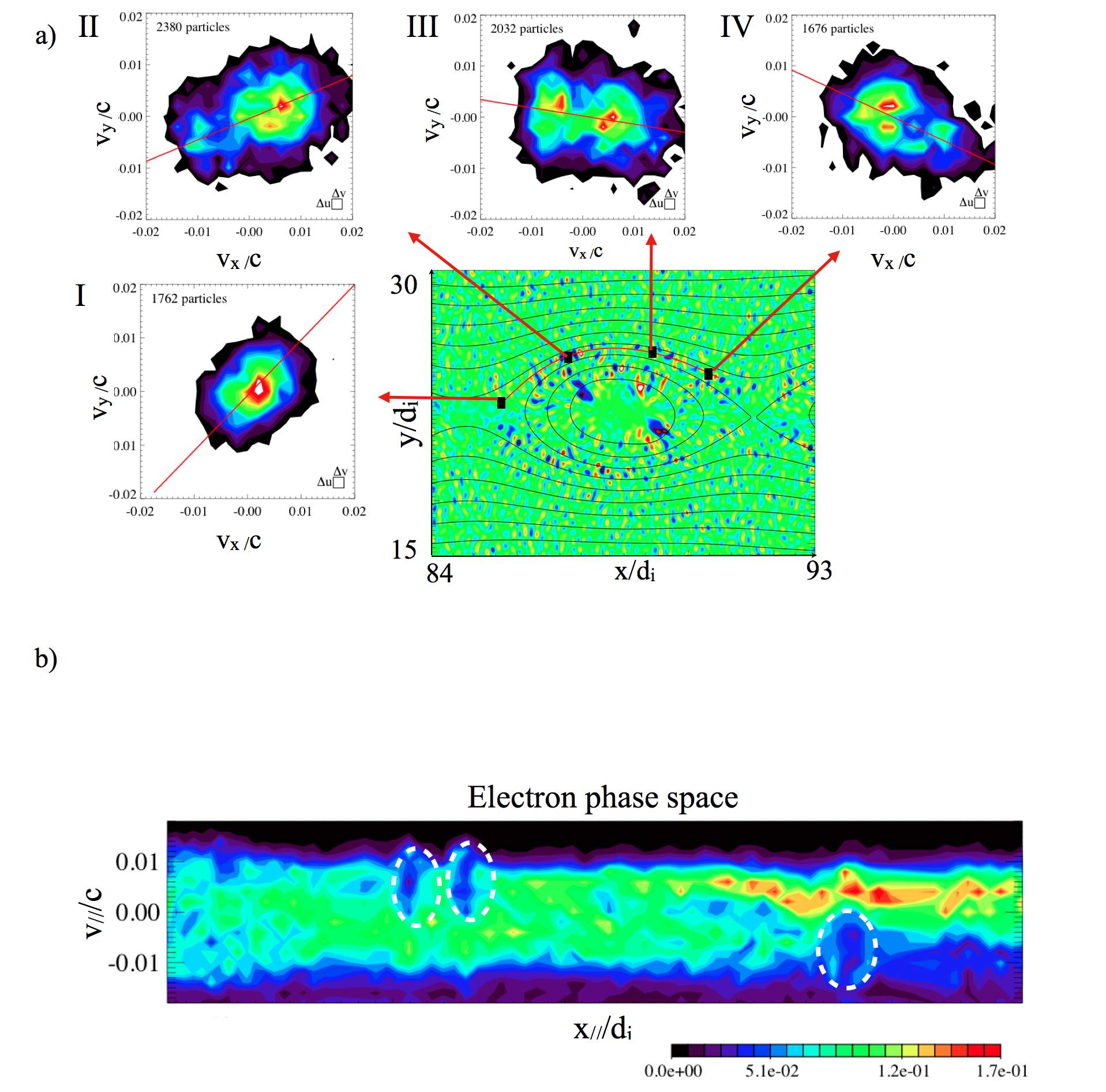}
\end{center}
\caption{Electron distribution functions, $x_{//}$ vs $v_{//}$ (curvilinear coordinate along the red line in panel a vs the parallel velocity), at four different locations (black boxes) at time $\Omega_{ci} t = 18.62$ in panel a. The red line in the distribution functions shows the direction of the magnetic field. Electron phase space along the red line of panel b with phase space holes, encircled in white dashed lines. The electron distribution functions show the presence of counter-streaming electron beams, while the phase space plots reveal the presence of electron phase pace holes along the separatrices. }
\end{figure*}

%\begin{figure*}[t]
%\vspace*{2mm}
%\begin{center}
%\includegraphics[width=16cm]{NPG_9.png}
%\end{center}
%\caption{Quiver plot of average ion velocity at time $\Omega_{ci} t = 10.48$ in the region $x=38d_i:47d_i,y=6.8d_y:13.5d_i$. The plot shows the ion trapping in the plasmoid.}
%\end{figure*}

%\begin{figure*}[t]
%\vspace*{2mm}
%\begin{center}
%\includegraphics[width=16cm]{NPG_10.png}
%\end{center}
%\caption{Ion distribution function $\Omega_{ci} t = 10.48$. Ion heating in the plasmoid is caused by the in-plane electric fields.}
%\end{figure*}

\section{Discussion}
%% \conclusions[modified heading if necessary]
Magnetic reconnection in a plasmoid chain has been studied by kinetic simulations. Magnetic reconnection is initially triggered by the tearing instability in multiple $X$~points. Several plasmoids originate from the outflow plasma of magnetic reconnection.  The reconnection rate calculated for different points reaches a maximum in a very short period $\Omega_{ci} t = 1.5$. After a very fast transient, the plasmoid chain dynamics is dominated by plasmoid coalescence. In this second phase, the reconnection rate decreases rapidly reaching an asymptotic values range $0.01 - 0.08$ for different $X$~points. 

Several high density regions, the plasmoids develop as a result of the initial tearing instability. Plasmoids grow in size by merging. At a given time, the plasmoid are approximately the same size. Secondary plasmoid formation and ejection of smaller plasmoids are not observed. The self-similar plasmoid hierarchy in the plasmoid chain \citep{Uzdensky:2010} is not present in the reported simulation. This is in contrast with previous plasmoid chain  Particle-in-Cell simulation results \citep{OkaAPJ:2010, OkaJGR:2010}, where secondary plasmoids were indeed detected. The difference between previous Particle-in-Cell simulations with the here presented results, might be explained by the use of different simulation box lengths, plasma temperatures, and ion to electron mass ratios. It is important to note that collisionless Particle-in-Cell simulations with realistic mass ratio do not report the formation of secondary plasmoids \citep{Lapenta:2010,LapentaGRL:2011}. Further studies are necessary to identify which simulation parameters determine the occurrence of secondary plasmoids in Particle-in-Cell simulations.

An out-of-plane core magnetic field develops in correspondence of the plasmoids. The strong core magnetic field has been observed several times in satellite observations. ISSE 3 and GEOTAIL missions reported the presence of cross-tail magnetic field in the Earth's magnetotail \citep{Slavin:1989}. Moreover, MHD \citep{Hesse:1996} and hybrid simulations \citep{Karimabadi:1999} confirmed the formation of a plasmoid core field and proposed a formation mechanisms in terms of Hall-generated currents. Moreover, hybrid simulations \citep{Karimabadi:1999} revealed that the core magnetic field has a complex structure originating from the complex flow pattern in the plasmoid. In multiple $X$~points reconnection hybrid simulation, S-shaped $B_z$ regions were found.  The fully kinetic simulation in this paper confirms the formation of unipolar core magnetic field and complex $B_z$ pattern. Some of these structures, such as S-shaped regions, can be explained as a result of the Hall magnetic field in anti-reconnection. 

The most intense electric fields develop in-plane, reaching a peak value that is approximately ten times the maximum reconnection electric field. They are localized along the separatrices: the electric fields point inward/(resp. outward) inside/(resp. outside) the plasmoid. Intense normal electric fields have been observed in magnetotail and an ion acceleration mechanism due by repetitive ion reflections between the separatrices have been proposed \citep{Wygant:2005}. The kinetic simulations in this paper shows an electric field configuration, that can support such acceleration mechanism \citep{Wygant:2005} and is agreement with the results of Particle-in-Cell studies \citep{MarkidisJGR:2011,MarkidisPOP:2011}. 
The kinetic study of plasmoid chain reveals that the separatrices are regions populated by counter-streaming electron beams, eventually driving the two-stream instability. An analysis of the electron distribution functions and phase space along the separatrix shows the presence of electron beams along the separatrices and the formation of electron phase space holes. Bipolar electric field structures were previously observed in \citep{Drake:2005}. In this study, it is proven that bipolar electric field structures are associated with electron phase space holes, produced by the two-stream instability of electron beams along the separatrices. The two-stream instability could generate the electron holes observed in plasmoid embedded in the magnetotail current sheet \citep{Khotyaintsev:2010}.

%\section{\\ \\ \hspace*{-7mm} HEADING}    %% Appendix A

%\subsection                               %% Appendix A1, A2, etc.

\conclusions
The general properties of magnetic reconnection in a plasmoid chain in the kinetic framework have been studied by analyzing two dimensional Particle-in-Cell simulation results. A plasmoid chain is rapidly formed by the tearing instability and plasmoids coalescence progressively in larger plasmoids. Secondary islands are not formed and plasmoids are approximately the same size with no hierarchical self-similar structure. This is probably a result of the chosen simulation parameters. Future work will investigate which quantities regulate the secondary plasmoid generation in Particle-in-Cell simulations. Each plasmoid is characterized by a core out-of-plane magnetic field and an out-of-plane electron current that drives the coalescence. The quadrupolar structure of the Hall magnetic field is the signature of anti-reconnection. Bipolar electric field structures are present along the plasmoid separatrices. The analysis of the electron distribution functions and phase space show counter-streaming electron beams, unstable to the two-stream instability, and electron phase space holes along the reconnection separatrices.

The plasmoid chain formation and evolution have been presented in two dimensional simulations.  The reduced geometry allows to simulate the plasmoid chains with realistic parameters in relative large systems (hundreds $d_i$), but the physical phenomena developing in the third direction are neglected. In real three dimensional systems, the magnetic islands correspond to extended flux ropes, interacting in a variety of complex ways non possible in two dimensional geometries. The comparison of two and three dimensional MHD simulation results show significant differences in the evolution of turbulent reconnection \citep{Kulpa:2010, Kowal:2009}. In the kinetic framework,  micro-instabilities in the third dimensions, such as the lower hybrid instabilities  \citep{Ricci:2005}, are not included in the model. In fact, recent three-dimensional simulations of magnetic reconnection with guide field \citep{Daughton:Turb1,Daughton:Turb2} show that oblique tearing modes can generate turbulence along the reconnection separatrices.  Future three dimensional studies are necessary to evaluate the effect of flux rope interaction and waves propagating in the third direction on the plasmoid chain evolution. 

\begin{acknowledgements}
The present work is supported by the NASA MMS Grant NNX08AO84G, by the Onderzoekfonds KU Leuven (Research Fund KU Leuven) and by the European Commission's Seventh Framework Programme (FP7/2007-2013) under the grant agreement no. 263340 (SWIFF project, www.swiff.eu). Simulations were conducted on the resources of the NASA Advanced Supercomputing Division (NAS), of the NASA Center for Computational Sciences Division (NCCS).
\end{acknowledgements}

\providecommand{\noopsort}[1]{}\providecommand{\singleletter}[1]{#1}%

\addtocounter{figure}{-1}\renewcommand{\thefigure}{\arabic{figure}a}

\end{document}